\documentclass[aip,pop,groupedaddress,reprint]{revtex4-2}
\usepackage{physics}
\usepackage{amsmath}
\usepackage{autobreak}
\usepackage{graphicx}
\usepackage{subfigure}
\usepackage{hyperref}
\usepackage{booktabs}
\usepackage{bm}

\DeclareMathOperator{\arctanh}{arctanh}
\DeclareMathOperator{\erfc}{erfc}

\begin{document}
\title{Investigations of the kinetic ion-acoustic soliton by the Bernstein-Greene-Kruskal integral method} 
\author{Ran Guo}
\thanks{Author to whom correspondence should be addressed}
\email{rguo@cauc.edu.cn}
\affiliation{Department of Physics, College of Science, Civil Aviation University of China, Tianjin 300300, China}
\begin{abstract}
The solitary waves are investigated through the Bernstein-Greene-Kruskal integral method with the ion response.
We consider two specific cases of ions, i.e., the single stream with the waterbag distribution and the two counter-propagating streams with the Maxwellian distribution.
The trapped electron distributions are derived for both two cases.
The results show that the trapped electron distribution can be either a hole or a hump in the phase space, depending on the competition between the contributions from the passing electron distribution, the potential profile, and the ion response.
We obtain the boundary between the ion-acoustic soliton and the electron hole in the parameter space.
The effects of the potential amplitude, width, and the ion-to-electron mass ratio on the separatrices are discussed.
The Vlasov simulations are conducted to verify the stability of the ion-acoustic soliton constructed by the integral method.
\end{abstract}
\maketitle

\section{Introduction}
\label{sec:intro}
The solitary wave with positive potential is a nonlinear structure commonly observed in space plasmas.
Matsumoto \textit{et al.} \cite{Matsumoto1994} identified the Broadband Electrostatic Noise in the plasma sheet boundary layer as a series of solitary waves.
Ergun \textit{et al.} \cite{Ergun1998} observed the solitary waves travelling much faster than the ion acoustic speed in the auroral zone.
Pickett \textit{et al.} \cite{Pickett2005} reported the observation of solitary waves throughtout the magnetosheath.
Graham \textit{et al.} \cite{Graham2016} investigated the speed, length scales, and field strengths of the solitary waves observed near the magnetopause and in the magnetosheath and found that the speeds of such solitary waves can range from almost stationary speed to the electron thermal speed in the ion frame.

The theoretical approaches to studying solitary waves are the phase space hole \cite{Schamel1972,Schamel1986,Hutchinson2017,Guo2022,Schamel2023, Hutchinson2024} and the soliton model \cite{Washimi1966,Sagdeev1966,Lakhina2018}.
On the one hand, the electron hole (EH) was studied by using the Bernstein-Greene-Kruskal (BGK) integral method and the Sagdeev pseudo-potential method.
The BGK method was first proposed in 1957 \cite{Bernstein1957} and then used to study the trapped electron distribution for several solitary potentials in the absence of the ion response. \cite{Turikov1984}
Krasovsky \textit{et al.} \cite{Krasovsky2003} developed the integral method with the ion response and studied the restrictions on the solitary-wave parameters and waveforms. 
Aravindakshan \textit{et al.} \cite{Aravindakshan2018} investigated the EH in suprathermal space plasmas and found that the suprathermal electrons can enlarge the parameter space allowed by the stable and weak EH.
On the other hand, the ion-acoustic soliton (IAS) was usually studied by the Sagdeev pseudo-potential method in the literature. \cite{Sagdeev1966}
Schamel proposed a widely used model representing the EH with the electron trapping parameter $\beta<0$ while the IAS with $\beta>0$. \cite{Schamel1972}
Jenab and Spanier conducted a series of kinetic simulations of the IAS modeled by the Schamel distribution with a wide range of trapping parameters. \cite{Jenab2016,Jenab2017,Jenab2017a,Jenab2018} 
Their results showed that the IAS can stably propagate even if the trapping parameter is positive and larger than one. \cite{Jenab2016}
However, the pseudo-potential method needs a pre-assumed distribution for the trapped electrons, which may exclude some possible characteristics of the solitary waves. \cite{Hutchinson2017}
Therefore, we expect to model the IAS by the BGK integral method so as to extend the allowable parameter space of the soliton.

In addition, the slow solitary wave, propagating with a speed close to or smaller than the ion-acoustic speed, attracts a wide range of interests.
These solitary waves were observed in the magnetotail\cite{Khotyaintsev2010,Lotekar2020} and magnetopause\cite{Graham2016}.
The early studies of electron hole (EH) indicated that if the ion response is considered, the slow EH would undergo self-acceleration and form a coupled hole-soliton (CHS). \cite{Saeki1998,Eliasson2004,Zhou2018} 
Then, Hutchinson \cite{Hutchinson2021a} proved that an electron phase space hole can be slow and stable
if the background ion distribution is double-humped and the speed of the solitary wave lies in the local minimum of the ion distribution.
This theory was supported by the observations. \cite{Kamaletdinov2021}
Therefore, the IAS with the double-humped ion distribution is also considered in this work.

The paper is organized as follows.
In Sec. \ref{sec:gef}, we introduce the plasma model and the general formulas for the trapped electron distribution with the ion response. 
In Sec. \ref{sec:one-WB-ion}, a single ion stream following the waterbag distribution is considered, the trapped electron distribution is derived, and the critical speed of the solitary wave between the IAS and the EH is obtained.
In Sec. \ref{sec:two-M-ions}, two counter-streaming ions following the Maxwellian distribution are supposed, the trapped electron distribution is numerically studied, and the separatrices between the IAS and the EH in the parameter space are discussed. 
In Sec. \ref{sec:sim}, we conduct numerical simulations to test the stability of the theoretical results.
In Sec. \ref{sec:summary}, the conclusions are made.

\section{Model and Theory}
\label{sec:model}
\subsection{General formulas}
\label{sec:gef}

We consider a one-dimensional electrostatic plasma system in the solitary wave frame.
For convenience, the dimensionless parameters are used in this study.
The length is scaled by the electron Debye length $\lambda_{De} = \sqrt{\epsilon_0 k_B T_e/(n_0 e^2)}$ and the velocity by the electron thermal speed $\sqrt{2 k_B T_e /m_e}$.
Here, $n_0$ and $T_e$ are, respectively, the undisturbed number density and temperature of electrons at $x\rightarrow \pm \infty$. 
The potential is measured in the unit of $k_B T_e/e$ and the energy in the unit of $k_B T_e$.
The number densities of electrons and ions are scaled by $n_0$.

The potential of the solitary wave is assumed to be positive, approaching its maximum $\psi$ at $x=0$ and vanishing at $x\rightarrow \pm \infty$.
Therefore, the electrons are divided into the passing and the trapped species.
The passing electron distribution is denoted by $f_{e,p}^{(+)}$ and $f_{e,p}^{(-)}$ for $v>0$ and $v<0$, respectively.
If the distribution of the passing electrons and the ions are given and the potential shape is known,
the trapped electron distribution can be solved by the BGK integral method, \cite{Bernstein1957,Turikov1984}
\begin{equation}
    f_{e,t}(w) = \frac{1}{\pi}\int_{0}^{-w} \frac{\dv*{n_{e,t}}{\phi}}{\sqrt{-w-\phi}} \dd{\phi},
    \label{eq:BGK_int}
\end{equation}
where $w = v^2 - \phi$ is the dimensionless electron energy, and $n_{e,t}$ is the number density of the trapped electrons.
Equation \eqref{eq:BGK_int} could be further simplified by employing the Poisson equation in dimensionless form,
\begin{equation}
    \dv[2]{\phi}{x} = n_e - n_i = n_{e,p}+n_{e,t}-n_i,  
    \label{eq:poisson}
\end{equation}
where the passing electron density is denoted by $n_{e,p}$ and the total one by $n_e = n_{e,p}+n_{e,t}$.
Differentiating the Poisson equation \eqref{eq:poisson} with respect to $\phi$ and rearranging the terms, one finds
\begin{equation}
    \dv{n_{e,t}}{\phi} = -\dv{n_{e,p}}{\phi} - \dv[2]{V}{\phi} + \dv{n_i}{\phi},
    \label{eq:dntdphi}
\end{equation} 
where $V(\phi)$ is the Sagdeev potential determined by the solitary potential shape,
\begin{equation}
    \dv{V}{\phi} = -\dv[2]{\phi}{x}.
    \label{eq:V}
\end{equation}
Substituting Eq. \eqref{eq:dntdphi} into Eq. \eqref{eq:BGK_int}, one can obtain the trapped electron distribution, \cite{Turikov1984,Krasovsky2003}
\begin{align}
    f_{e,t}       &= f_{e,t}^{(1)}+f_{e,t}^{(2)}+f_{e,t}^{(3)}, \label{eq:ft-general}  \\ 
    f_{e,t}^{(1)} &= \frac{\sqrt{-w}}{\pi} \int_0^\infty \frac{f_{e,p}^{(+)}(w')+f_{e,p}^{(-)}(w')}{2\sqrt{w'}(w'-w)} \dd{w'}, \label{eq:ft1_general}  \\
    f_{e,t}^{(2)} &= -\frac{1}{\pi} \int_0^{-w} \dv[2]{V}{\phi} \frac{1}{\sqrt{-w-\phi}}\dd{\phi}, \label{eq:ft2-general}\\
    f_{e,t}^{(3)} &= \frac{1}{\pi} \int_0^{-w} \dv{n_i}{\phi} \frac{1}{\sqrt{-w-\phi}}\dd{\phi}. \label{eq:ft3-general}
\end{align}
The first and second terms $f_{e,t}^{(1)}$, $f_{e,t}^{(2)}$ denote the contribution of the passing electron and the potential shape.\cite{Turikov1984}
The third term $f_{e,t}^{(3)}$ can be attributed to the ion response because it vanishes if the ion response is neglected by setting $\dv*{n_i}{\phi}=0$.

\subsection{Single ion stream with the waterbag distribution}
\label{sec:one-WB-ion}
\begin{figure}[tb]
    \centering
    \includegraphics[width=8.5cm]{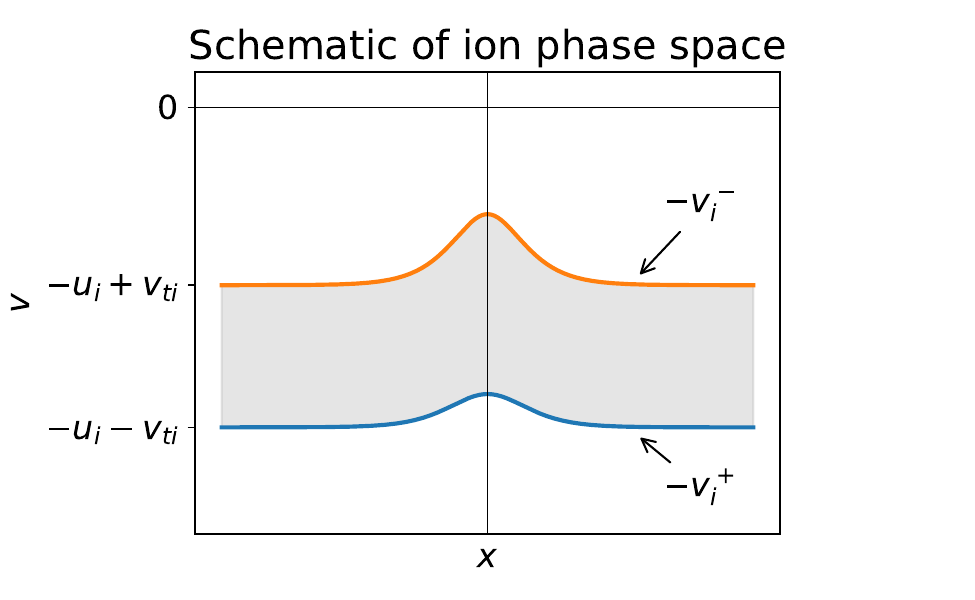}
    \caption{Schematic of the ion phase space. The ion distribution is a single-stream WB distribution, namely $f_i = 1/(2 v_{ti})$ in the gray shadows and $f_i = 0$ in the other regions.}
    \label{fig:ion-phase-space}
\end{figure}

The electrons are assumed to be Maxwellian and stationary relative to the solitary wave,
so the distribution for the passing electrons is,
\begin{equation}
    f_{e,p} = \frac{1}{\sqrt{\pi}} \exp(-v^2+\phi).
    \label{eq:fp}
\end{equation}
Therefore, $f_{e,t}^{(1)}$ can be calculated analytically, \cite{Turikov1984}
\begin{equation}
    f_{e,t}^{(1)} = \frac{1}{\sqrt{\pi}} e^{-w}\erfc(\sqrt{-w}),
    \label{eq:ft1}
\end{equation}
where $\erfc(z) = (2/\sqrt{\pi}) \int_z^\infty \exp(-t^2) \dd{t}$ is the complementary error function. \cite{Olver2010}
The solitary potential is chosen as,
\begin{equation}
    \phi = \psi \sech^2\left(\frac{x}{\Delta}\right),
    \label{eq:phi}
\end{equation}
to stand for the IAS with its amplitude $\psi$ and width $\Delta$.
It is worth noting that the present approach is a fully kinetic treatment; therefore, $\psi$ and $\Delta$ do not necessarily obey the well-known relationship between them for the IAS in the classic fluid model.
By inserting Eq. \eqref{eq:phi} into Eq. \eqref{eq:ft2-general}, the second term $f_{e,t}^{(2)}$ can be derived as, \cite{Turikov1984}
\begin{equation}
    f_{e,t}^{(2)} = \frac{8\sqrt{-w}}{\pi \Delta^2}\left(\frac{2w}{\psi}+1\right).
    \label{eq:ft2}
\end{equation}
The ions are assumed to follow the waterbag (WB) distribution, i.e., 
\begin{equation}
    f_{i} = \left\{
        \begin{aligned}
            &\frac{1}{2 v_{ti}}, \quad &\text{if } -u_i-v_{ti}<v<-u_i+v_{ti},\\
            &0, \quad &\text{otherwise},       
        \end{aligned}
    \right.
    \label{eq:fi}
\end{equation}
where $u_i$ is the bulk speed of ions, and $v_{ti}$ is the ion thermal speed.
In the dimensionless form, the energy conservations of the ions are,
\begin{equation}
    \mu (-u_i \pm v_{ti})^2 = \mu (v_i^{\pm})^2 + \phi,
    \label{eq:ion-energy-conservation}
\end{equation}
where $\mu = m_i/m_e$ is the ion-to-electron mass ratio.
We assume there are no reflected ions in this subsection, resulting in a lower limit of the ion drift speed
\begin{equation}
    u_i>u_r = v_{ti}+\sqrt{\psi/\mu}.
    \label{eq:ur}
\end{equation}
Figure \ref{fig:ion-phase-space} plots a schematic diagram of the WB distribution in ion phase space.
Therefore, one obtains the ion number density,
\begin{align}
    n_i(\phi) =& \frac{v_i^+ - v_i^-}{2 v_{ti}} \notag \\ 
        =& \frac{1}{2} \left[\sqrt{\left(\frac{u_i}{v_{ti}}+1\right)^2-\frac{\phi}{\mu v_{ti}^2}} - \sqrt{\left(\frac{u_i}{v_{ti}}-1\right)^2-\frac{\phi}{\mu v_{ti}^2}}\right],
    \label{eq:ni}
\end{align}
resulting in
\begin{multline}
    \dv{n_i}{\phi} = \frac{1}{4\mu v_{ti}^2} \left\{\left[\left(\frac{u_i}{v_{ti}}+1\right)^2-\frac{\phi}{\mu v_{ti}^2}\right]^{-\frac{1}{2}} \right. \\
    \left. -\left[\left(\frac{u_i}{v_{ti}}-1\right)^2-\frac{\phi}{\mu v_{ti}^2}\right]^{-\frac{1}{2}}\right\}.
    \label{eq:dni}
\end{multline}
Eventually, the third term $f_{e,t}^{(3)}$ can be calculated by substituting Eq. \eqref{eq:dni} into Eq. \eqref{eq:ft3-general},
\begin{align}
    f_{e,t}^{(3)} =& \frac{1}{2\sqrt{\mu} v_{ti} \pi}\left[\arctanh\left(\frac{\sqrt{-w/\mu}}{u_i-v_{ti}}\right)\right. \notag \\
    &\qquad \qquad \left. \qquad -\arctanh\left(\frac{\sqrt{-w/\mu}}{u_i+v_{ti}}\right)\right] \notag \\
    =& \frac{1}{2\sqrt{\mu} v_{ti} \pi} \arctanh \left(\frac{2v_{ti}\sqrt{-w/\mu}}{u_i^2-v_{ti}^2+w/\mu}\right).
    \label{eq:ft3-os}
\end{align}
The inverse hyperbolic function $\arctanh x$ requires $-1<x<1$, leading to $\mu(u_i-v_{ti})^2>\psi$ because of $-\psi<w<0$ for the trapped species.
This requirement is exactly the assumption of no reflected ions \eqref{eq:ur}.
The total distribution for the trapped electrons is,
\begin{multline}
    f_{e,t} = \frac{e^{-w}}{\sqrt{\pi}} \erfc(\sqrt{-w}) + \frac{8\sqrt{-w}}{\pi \Delta^2}\left(\frac{2w}{\psi}+1\right) \\
    + \frac{1}{2\sqrt{\mu} v_{ti} \pi} \arctanh \left(\frac{2v_{ti}\sqrt{-w/\mu}}{u_i^2-v_{ti}^2+w/\mu}\right).
    \label{eq:ft}
\end{multline}
We have known that the sum of the first two terms in Eq. \eqref{eq:ft}, i.e., $f_{e,t}^{(1)}+f_{e,t}^{(2)}$, has the minimum value at $w=-\psi$, resulting in a hole in the electron phase space in the case of no ion response. \cite{Turikov1984}
However, the third term $f_{e,t}^{(3)}$ is a monotonic decreasing function of $w$, leading to $f_{e,t}^{(3)}(w=-\psi) > f_{e,t}^{(3)}(w=0)$.
Therefore, the total trapped electron distribution $f_{e,t}$ can be either a hole or a hump in the phase space, depending on the competition between $f_{e,t}^{(1)}+f_{e,t}^{(2)}$ and $f_{e,t}^{(3)}$.

We use the condition whether there is a hole or a hump in phase space to distinguish between the EH and the IAS.
Therefore, the critical condition between the soliton and the EH could be given by,
\begin{equation}
    f_{e,t}(w=-\psi) = f_{e,t}(w=0),
    \label{eq:cond-sep}
\end{equation}
which is equivalent to,
\begin{multline}
    e^{\psi}\erfc(\sqrt{\psi}) - 1 - \frac{8\sqrt{\psi}}{\sqrt{\pi} \Delta^2} \\
    + \frac{1}{2\sqrt{\mu\pi} v_{ti}} \arctanh \left(\frac{2v_{ti}\sqrt{\psi/\mu}}{u_c^2-v_{ti}^2+\psi/\mu}\right) = 0.
    \label{eq:sep-s-eh}
\end{multline}
Here, $u_{c}$ is the critical drift speed, which can be solved from the above equation \eqref{eq:sep-s-eh} for the given $\psi$, $\Delta$, $v_{ti}$, and $\mu$,
\begin{multline}
    u_{c}^2 = v_{ti}^2 + \frac{\psi}{\mu} \\ 
    + \frac{2v_{ti}\sqrt{\psi/\mu}}{\tanh\left\{2v_{ti}\sqrt{\mu\pi}\left[1-e^\psi\erfc{\sqrt{\psi}}+(8/\Delta^2)\sqrt{\psi/\pi}\right]\right\}}.
    \label{eq:uic}
\end{multline}
Due to the range of the hyperbolic tangent function $-1<\tanh x<1$,
it is straightforward that $u_c$ is always larger than the lower limit of ion drift speed $u_r$ \eqref{eq:ur}
given by the assumption of no reflected ions. 
Therefore, the solitary wave is an IAS with its speed in the range of $u_r<u_i<u_c$, while it is an EH when $u_i>u_{c}$.

\begin{figure}[tb]
    \centering
    \includegraphics[width=8.5cm]{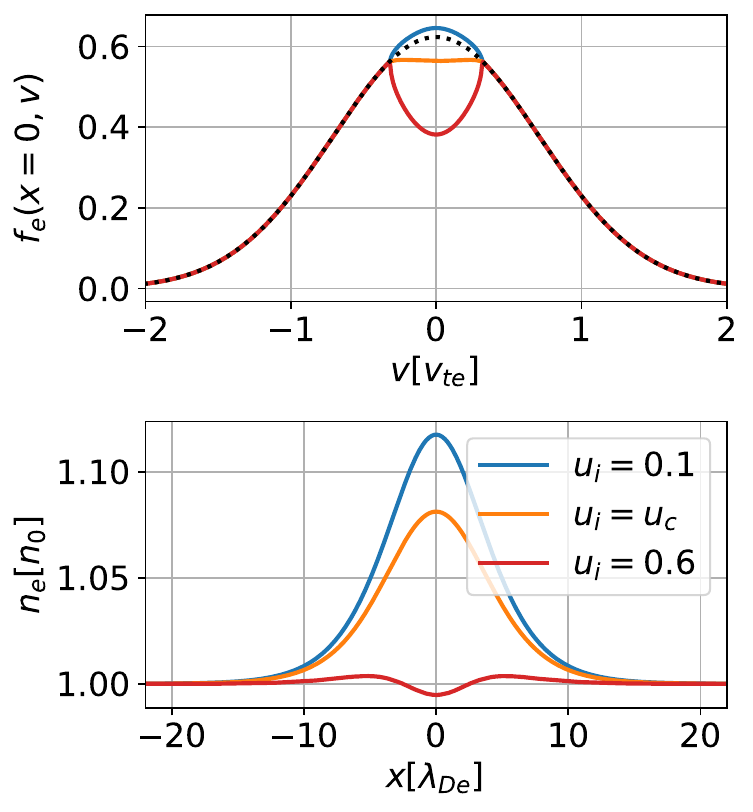}
    \caption{
        The electron velocity distribution at $\phi = \psi$ (upper panel) and the electron density (lower panel) for the IAS and EH models, respectively.
        The potential amplitude is assumed to be $\psi = 0.1$, and the width is $\Delta = 5$.
        The ion thermal speed $v_{ti} = 0.02$ is set, and the ion-to-electron mass ratio $\mu = 50$.
        The critical speed is $u_c \approx 0.115$ calculated by Eq. \eqref{eq:uic}.
        In the upper panel, the black dotted line is the referenced Maxwellian distribution $f_M = \exp(-v^2+\psi)/\sqrt{\pi}$.
    }
    \label{fig:fe}
\end{figure}
Figure \ref{fig:fe} shows the electron distribution calculated by Eqs. \eqref{eq:fp} for the passing species and \eqref{eq:ft} for the trapped ones at $\phi = \psi$.
It shows that the trapped electron distribution is a hump for $u_i < u_{c}$, an approximate flat top for $u_i = u_{c}$ in this case, and a hole for $u_i > u_{c}$.
In addition, it is proved (in Appendix \ref{app:consistency}) that the trapped electron distribution is Maxwellian if one adopts the exact solution of the pure soliton from the fluid theory.

The critical speed $u_c$ \eqref{eq:uic} can be equivalently expressed as,
\begin{multline}
    \frac{u_{c}^2}{c_s^2} = \frac{v_{ti}^2}{c_s^2} + 2\psi \\ 
    + \frac{2\frac{v_{ti}}{c_s}\sqrt{2\psi}}{\tanh\left\{\frac{v_{ti}}{c_s}\sqrt{2\pi}\left[1-e^\psi\erfc{\sqrt{\psi}}+\frac{8}{\Delta^2}\sqrt{\frac{\psi}{\pi}}\right]\right\}},
    \label{eq:uic2cs}
\end{multline}
where $c_s$ is the ion-acoustic speed given by $c_s = \flatfrac{\sqrt{T_e/m_i}}{\sqrt{2T_e/m_e}} = \sqrt{1/(2\mu)}$ in the unit of $v_{te}$.
In this form, the critical speed $u_{c}/c_s$ does not explicitly involve the ion-to-electron mass ratio $\mu$.
\begin{figure}[tb]
    \centering
    \includegraphics[width=8.5cm]{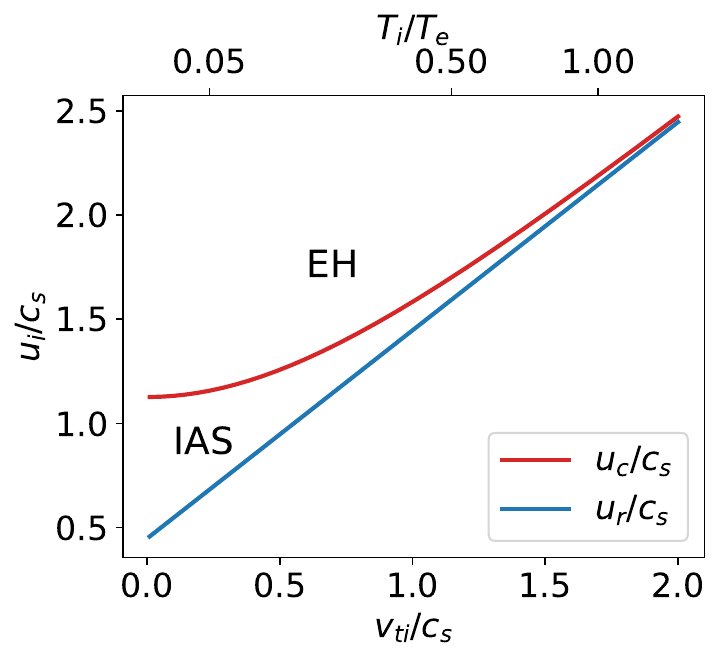}
    \caption{
            Parameter space of the ion drift speed $u_i/c_s$ versus the thermal speed $v_{ti}/c_s$ (lower $x$-axis) and ion-to-electron temperature ratio $T_i/T_e$ (upper $x$-axis) for the solitary wave.
            The relationship between the ion thermal speed and the ion temperature in the waterbag model is $v_{ti} = \sqrt{(3/2) (T_i/\mu)}$ in the dimensionless form.
            The red line illustrates the critical drift speed $u_{c}/c_s$ \eqref{eq:uic2cs} separating the parameter spaces of the IAS and the EH.
            The blue line is the lower limit of the ion drift speed $u_r/c_s$ \eqref{eq:ur}.
            The other parameters are selected as $\psi = 0.1$ and $\Delta = 5$.
    }
    \label{fig:uc}
\end{figure}
Figure \ref{fig:uc} illustrates the critical drift speed $u_c/c_s$ dividing the parameter space of $u_i/c_s$ and $v_{ti}/c_s$ into the IAS and EH regions.
It shows that the critical speed $u_c$ approaches the lower limit $u_r$ \eqref{eq:ur} for a large $v_{ti}$. 
The reason is that the hyperbolic tangent function in the denominator of the third term of Eq. \eqref{eq:uic2cs} approaches one when its argument is large, leading to $u_c/c_s \rightarrow v_{ti}/c_s+\sqrt{2\psi} = u_r/c_s$.
In addition, an increased $\psi$ and a decreased $\Delta$ can also enhance the argument of the hyperbolic tangent function in Eq. \eqref{eq:uic2cs}, resulting in the same effect on the critical speed $u_c$ and thus suppressing the IAS region in the parameter space.

In Fig. \ref{fig:uc}, it seems that the IAS can propagate slower than the ion-acoustic speed with a single cold ion stream $T_i \ll T_e$ and non-Maxwellian electrons.
However, in this case, the plasma may be unstable due to the ion-ion instabilities in terms of the linear theory. \cite{Krall1973}
The stability of the IAS constructed by the BGK method would be tested by the Vlasov simulations in Sec. \ref{sec:sim}.

In addition, when $u_i$ takes the value close to the critical speed $u_c$, the trapped electron distribution exhibits an interesting feature, i.e., a hump (hole) but with a local minimum (maximum) at the center $x=0, v=0$.
This behavior could be described by the second partial derivative of $f_{e,t}$ with respect to $v$ at the hump/hole center.
After differentiating Eq. \eqref{eq:ft}, one has
\begin{multline}
    \pdv[2]{f_{e,t}}{v}\Big\rvert_{x=0,v=0} = \frac{2}{\pi\sqrt{\psi}}  - \frac{2e^\psi}{\sqrt{\pi}} \erfc(\sqrt{\psi}) 
                            + \frac{40}{\pi\Delta^2 \sqrt{\psi}} \\
                          - \frac{1}{\mu\pi\sqrt{\psi}} \frac{u_i^2-v_{ti}^2+\psi/\mu}{(u_i^2-v_{ti}^2-\psi/\mu)^2-4v_{ti}^2\psi/\mu}.
    \label{eq:d2f-os}
\end{multline}  
The first two terms on the right-hand side of Eq. \eqref{eq:d2f-os} are the contributions from the passing electrons, the third term from the potential shape, and the last one from the ion response.
It could be proved that the sum of the first three terms is positive (in Appendix \ref{app:sign}), leading to the hole center being a local minimum without the ion response.
However, when the ion response is involved, 
the center of the trapped electron distribution could be either a local maximum or minimum, whether it is an EH or IAS.
We use the notation $u_c^*$ to denote the speed at which the above second derivative \eqref{eq:d2f-os} vanishes, and then one could solve two solutions of $u_c^*$.
Neglecting the solution less than the lower limit $u_r$ (see Appendix \ref{app:sol-ucs}), we derive $u_c^*$ as,
\begin{equation}
    u_c^{*2} = v_{ti}^2 + \frac{\psi}{\mu} + \frac{1}{4 \mu A} 
                + \sqrt{\frac{1}{16 \mu^2 A^2} + \frac{\psi}{\mu^2 A} + 4 v_{ti}^2 \frac{\psi}{\mu}},
    \label{eq:uc-star}
\end{equation}
or equivalently,
\begin{equation}
    \frac{u_c^{*2}}{c_s^2} = \frac{v_{ti}^2}{c_s^2} + 2\psi + \frac{1}{2 A} 
                + \sqrt{\frac{1}{4 A^2} + \frac{4\psi}{A} + 8\frac{v_{ti}^2}{c_s^2} \psi},
    \label{eq:uc-star2cs}
\end{equation}
where $A = 1-\sqrt{\pi \psi}\exp(\psi)\erfc(\sqrt{\psi})+20/\Delta^2$.
\begin{figure}[tb]
    \centering
    \includegraphics[width=8.5cm]{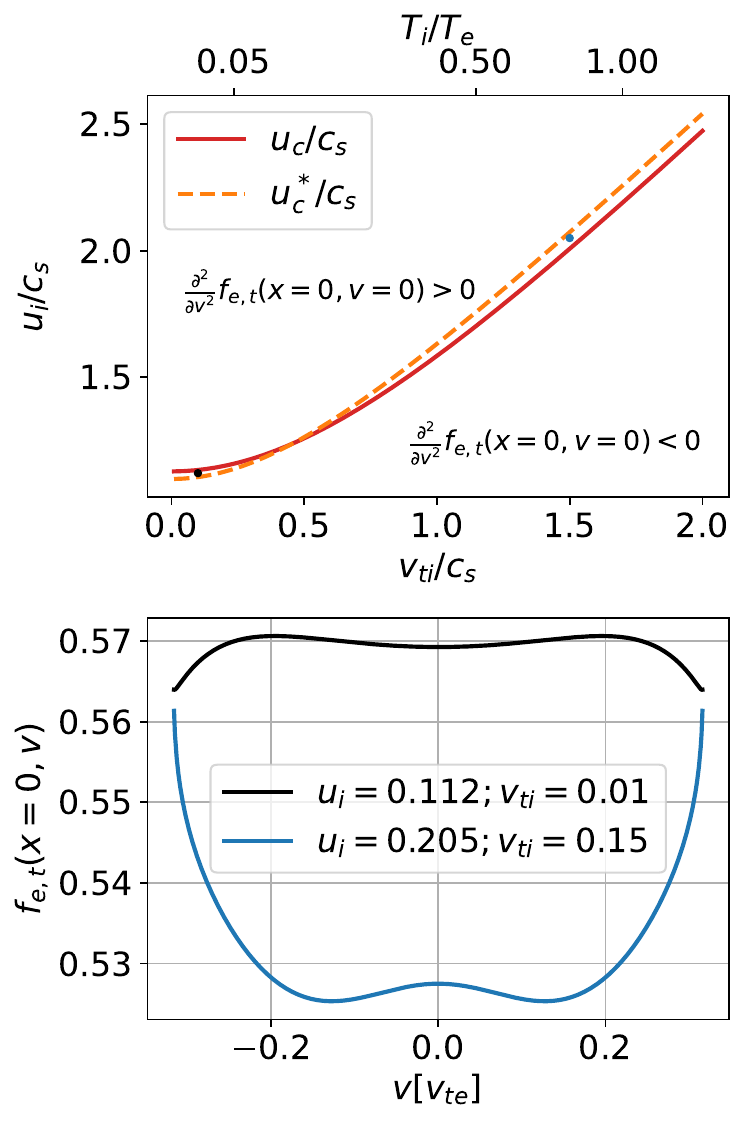}
    \caption{
        The separatrix where the second partial derivative $\pdv*[2]{f_e}{v} = 0$ at $x=0, v=0$ (upper panel) and the trapped electron distribution $f_{e,t}(x=0,v)$ (lower panel).
        In the upper panel, the solid red line draws the critical speed $u_c/c_s$ \eqref{eq:uic2cs} as the same red line in Fig. \ref{fig:uc}.
        The orange dashed line plots the speed $u_c^*/c_s$ \eqref{eq:uc-star2cs} leading to $\pdv*[2]{f_e}{v} = 0$ at $x=0, v=0$.
        The second partial derivative is positive (negative) for $u_i>u_c^*$ ($u_i<u_c^*$).
        The lower panel shows the trapped electron distribution for two sets of $u_i$ and $v_{ti}$.
        The distribution $f_{e,t}$ plotted by the black (blue) line adopts the parameters marked by the black (blue) dot in the upper panel.
        The other parameters are $\psi = 0.1$, $\Delta = 5$, and $\mu = 50$.
    }
    \label{fig:uc-star}
\end{figure}
In Fig. \ref{fig:uc-star},
the upper panel illustrates the comparison between the critical speeds $u_c$ and $u_c^*$.  
Although the curves representing $u_c$ and $u_c^*$ are very close to each other, there is still a narrow gap between them.
We purposely choose two sets of $u_i$ and $v_{ti}$ in the gap between $u_c$ and $u_c^*$ (marked by black and blue point in the upper panel)
and plot the trapped electron distributions $f_{e,t}(x=0,v)$ with such parameters in the lower panel of Fig. \ref{fig:uc-star}.
The results show the possibility that an EH may have a local maximum at the center $x=0, v=0$, while an IAS may have a local minimum.
It should be stressed that the second partial derivative $\pdv*[2]{f_e}{v}$ only describes the local feature of the distribution at the center $x=0, v=0$, so we still use the condition \eqref{eq:cond-sep} to distinguish between the IAS and EH in this work.

\subsection{Two ion streams with the Maxwellian distribution}
\label{sec:two-M-ions}
The one-dimensional Vlasov equation requires that the stationary distribution should be the function of the energy.
\begin{figure}[tb]
    \centering
    \includegraphics[width=8.5cm]{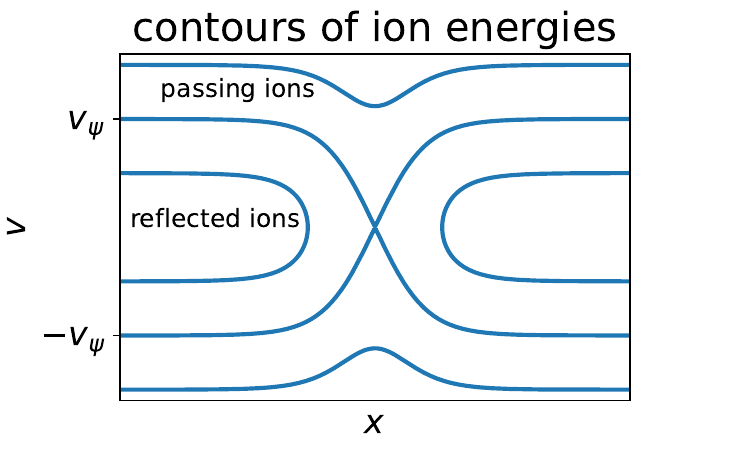}
    \caption{The energy contours of ions for a positive solitary potential in the phase space. }
    \label{fig:energy-contours}
\end{figure}
So, the undisturbed distribution of ions should be symmetric between the speeds $-v_\psi$ and $v_\psi$, where $v_\psi = \sqrt{\psi/\mu}$ is the speed separating the passing and reflected ions at $x=\pm\infty$, as shown in Fig. \ref{fig:energy-contours}.
For the case of the single-stream Maxwellian ions, the only possibility is the non-drifting distribution, i.e., $u_i = 0$.
However, it was proved that the EH is unstable in such a case and would undergo self-acceleration. \cite{Eliasson2004,Hutchinson2021a}
Previously, Hutchinson showed that a slow and stable EH can be constructed if the ions follow a double-humped distribution. \cite{Hutchinson2021a}
Therefore, our interest is to study whether such a slow solitary wave becomes a soliton if the ion drift speed is small enough.
Here, the soliton refers to the trapped electron distribution being a hump in the phase space.
Therefore, we consider the Maxwellian distribution for two counter-streaming ions,
\begin{multline}
    f_{i} = \frac{1}{2\sqrt{\pi} v_{ti}} \left\{\exp\left[-\frac{(\sqrt{v^2+\phi/\mu}+u_i)^2}{v_{ti}^2}\right] \right. \\
    \left. +\exp\left[-\frac{(\sqrt{v^2+\phi/\mu}-u_i)^2}{v_{ti}^2}\right]\right\},
    \label{eq:fi2}
\end{multline}
which yields the ion number density,
\begin{multline}
    n_i(\phi) = \frac{1}{\sqrt{\pi}v_{ti}} \int_0^\infty \dd{v} \left\{
        \exp\left[-\frac{(\sqrt{v^2+\phi/\mu}+u_i)^2}{v_{ti}^2}\right] \right. \\
        \left. +\exp\left[-\frac{(\sqrt{v^2+\phi/\mu}-u_i)^2}{v_{ti}^2}\right] \right\}
    \label{eq:ni2}
\end{multline}
due to the symmetry of the ion velocity distribution.
\begin{figure}[tb]
    \centering
    \includegraphics[width=8.5cm]{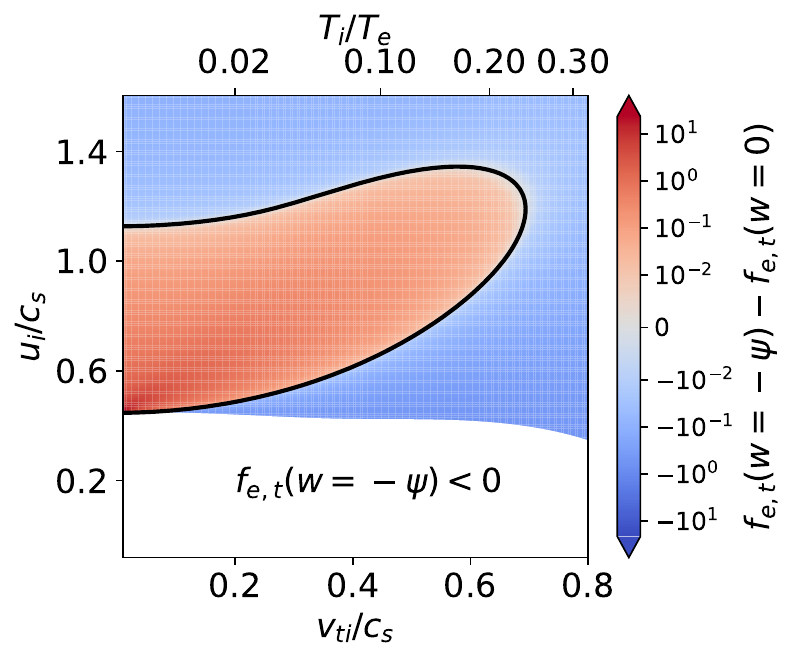}
    \caption{The difference $f_{e,t}(w=-\psi)-f_{e,t}(w=0)$ in the case of the two ion streams.
        The potential amplitude $\psi = 0.1$ and the width $\Delta = 5$ are assumed.
        The black line is $f_{e,t}(-\psi)-f_{e,t}(0)=0$ which splits the IAS (red region) and EH (blue region).
        In the blank area, the trapped electron distribution is negative at $w=-\psi$, which is unavailable in physics.
    }
    \label{fig:ft_psi_sub_zero}
\end{figure}
Substituting Eq. \eqref{eq:ni2} into Eq. \eqref{eq:ft3-general}, one derives the ion contribution of the trapped electron distribution $f_{e,t}^{(3)}$,
\begin{equation}
    f_{e,t}^{(3)} = -\frac{1}{\pi^{3/2}\mu v_{ti}^3} \int_0^{-w} \dd{\phi} \frac{I(\phi)}{\sqrt{-w-\phi}}
    \label{eq:ft3-ts}
\end{equation}
with the integral $I(\phi)$ defined by,
\begin{multline}
    I(\phi) = \\ 
    \int_0^\infty \dd{v} \Bigg\{ \exp\left[-\frac{(\sqrt{v^2+\phi/\mu}+u_i)^2}{v_{ti}^2}\right]
        \left(1+\frac{u_i}{\sqrt{v^2+\phi/\mu}}\right) \\
        +\exp\left[-\frac{(\sqrt{v^2+\phi/\mu}-u_i)^2}{v_{ti}^2}\right]
        \left(1-\frac{u_i}{\sqrt{v^2+\phi/\mu}}\right)
    \Bigg\}. 
    \label{eq:I}
\end{multline}
Eqs. \eqref{eq:ft3-ts} and \eqref{eq:I} are rather complicated expressions, so we numerically study their properties.
We still use the condition \eqref{eq:cond-sep} to distinguish between the IAS and EH.
Figure \ref{fig:ft_psi_sub_zero} illustrates the difference between the trapped electron distributions at the hole/hump center $w = -\psi$ and the edge $w = 0$.
The positive value of $f_{e,t}(w=-\psi)-f_{e,t}(w=0)$ (red region) indicates a hump in the phase space, while the negative value (blue region) indicates a hole.
In the case of the low thermal speed $v_{ti}$, the reflected ions play a less important role, so the trapped electron distribution behaves like that for WB ions discussed in the previous Sec. \ref{sec:one-WB-ion}.
During the increased thermal speed $v_{ti}$, the population of reflected ions grows, and that of the passing ones decreases, reducing the ion density at the peak of the solitary potential.
When such a reduction is substantial enough, the ion density perturbation cannot support the solitary wave.
So, the electron density deficit, i.e., the EH in the phase space, must exist to maintain the positive potential.
Therefore, the IAS cannot exist with a sufficiently large $v_{ti}$, as shown in Fig. \ref{fig:ft_psi_sub_zero}.
\begin{figure}[tb]
    \centering
    \includegraphics[width=8.5cm]{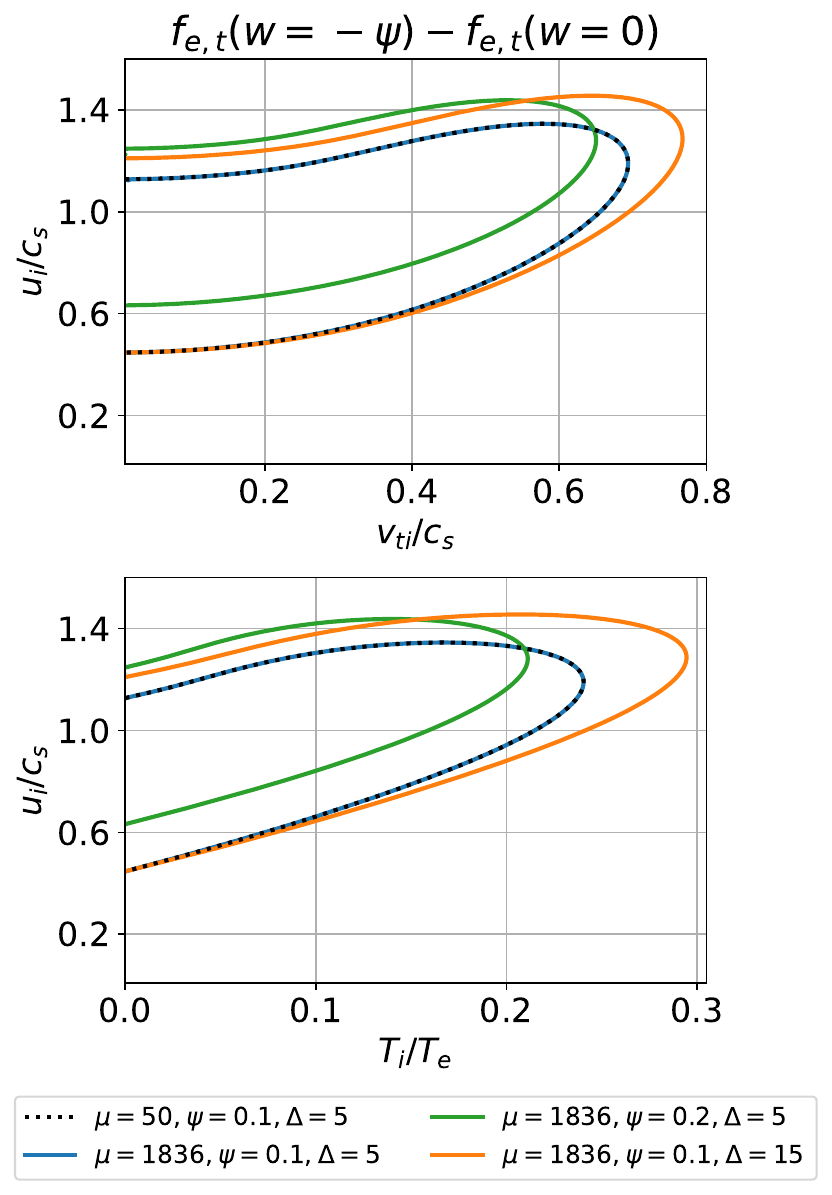}
    \caption{The separatrices of the IAS and EH for different parameters $\mu$, $\psi$, and $\Delta$.
    The separatrices are plotted in the upper panel for $u_i/c_s$ versus $v_{ti}/c_s$, while in the lower panel for $u_i/c_s$ versus $T_i/T_e$.
    }
    \label{fig:sep_s_eh}
\end{figure}
Figure \ref{fig:sep_s_eh} draws the separatrices between the IAS and EH for different parameters.
According to the typical parameters of the observations \cite{Lotekar2020} $\psi \sim 0.1$ and $\Delta \sim 10$,
the IAS can only survive in the case of $T_i/T_e \lesssim 0.3$ with two counter-streaming Maxwellian ions.
Different ion-to-electron mass ratios $\mu$ do not affect the separatrix due to the choice of the coordinate scale, which is consistent with the case of a single ion stream with the WB distribution, i.e., Eq. \eqref{eq:uic2cs}.

For the IAS with two counter-streaming ions, the trapped electron distribution $f_{e,t}$ is a hump overall in the phase space but may have a local minimum at the center $x=0, v=0$, similar to the case of the single ion stream.
The second partial derivative $\pdv*[2]{f_{e,t}}{v}$ at the center $x=0, v=0$ can be calculated by, 
\begin{align}
    \pdv[2]{f_{e,t}}{v}\Big\rvert_{x=0,v=0} = & \frac{2}{\pi\sqrt{\psi}}  - \frac{2e^\psi}{\sqrt{\pi}} \erfc(\sqrt{\psi}) 
                            + \frac{40}{\pi\Delta^2 \sqrt{\psi}} \notag \\
                          & + \frac{2}{\pi^{3/2}\mu v_{ti}^3} \left(\frac{I(0)}{\sqrt{\psi}}+\int_0^\psi \dd{\phi}\frac{\dv*{I}{\phi}}{\sqrt{\psi-\phi}}\right),
    \label{eq:d2f-ts}
\end{align}  
where the integral $I(\phi)$ is defined in Eq. \eqref{eq:I} (see details in Appendix \ref{app:dev-d2f}).
\begin{figure}[tb]
    \centering
    \includegraphics[width=8.5cm]{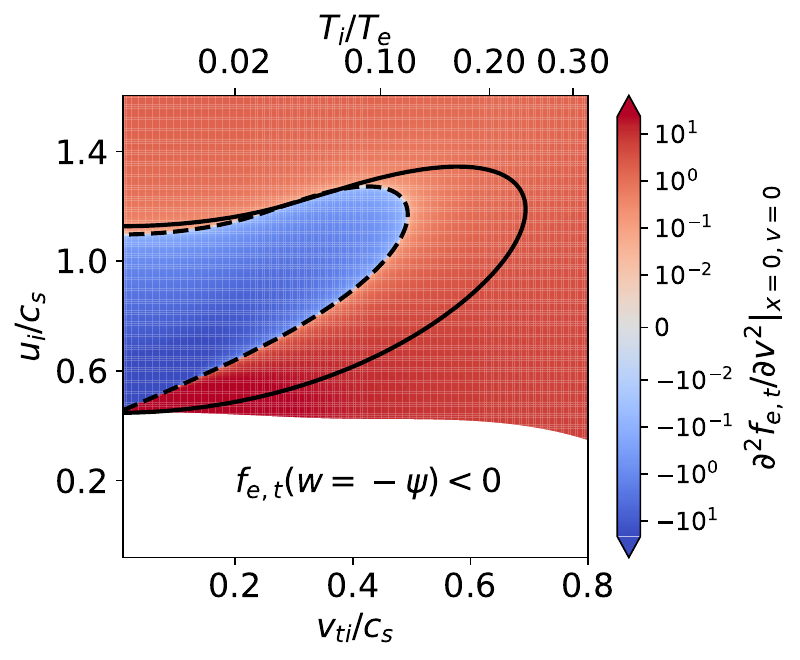}
    \caption{The second partial derivative $\pdv*[2]{f_{e,t}}{v}$ at the center $x=0, v=0$ for the two ion streams.
        The solid black line is the boundary between the IAS and EH, which is the same line in Fig. \ref{fig:ft_psi_sub_zero}.
        The red (blue) region represents the positive (negative) second derivative, indicating a local minimum (maximum).
        The other parameters are $\psi = 0.1$ and $\Delta = 5$.
    }
    \label{fig:d2f-ts}
\end{figure}
Figure \ref{fig:d2f-ts} shows the numerical results of $\pdv*[2]{f_{e,t}}{v}$ at the center $x=0, v=0$.
It illustrates that the IAS may have a local minimum (red region inside the black solid line) at the center $x=0, v=0$, while the EH always has a local minimum (red region outside the black solid line).

\section{The tests of stability by simulations}
\label{sec:sim}
As the IAS model constructed in this work is a self-consistent solution of the Vlasov-Poisson system, it does not evolve immediately on a fast timescale. 
However, it might experience instabilities in long-time simulations, such as the hole acceleration. \cite{Eliasson2004,Hutchinson2021a}
To test the IAS stability, we perform the one-dimensional Vlasov-Poisson simulations,
which solve the Vlasov equation by the semi-Lagrangian splitting scheme with cubic spline interpolations, \cite{Cheng1976}
and the Poisson equation by the tridiagonal matrix algorithm. \cite{Press2007}
The codes were verified by our previous simulations of linear Landau damping for the electron-acoustic waves in Kappa-distributed plasmas. \cite{Guo2021a}
In the present work, the simulation is conducted in the phase space domain $[-L/2, L/2] \times [-v_{max},v_{max}]$ with $L=80 \lambda_{De}$ and $v_{max}= 5 v_{te}$.
The open boundary conditions are adopted in the position space,
and the distribution is treated as zero when the velocity is outside the interval ($|v|>v_{max}$).
The spatial domain is discretized by $N_x = 1000$ grid points and the velocity space by $N_v=2000$ grid points.
The time step is $\dd t=0.005 \omega_{pe}^{-1}$, and the simulation time is $t=500 \omega_{pe}^{-1}$.
\begin{figure}[tb]
    \centering
    \includegraphics[width=8.5cm]{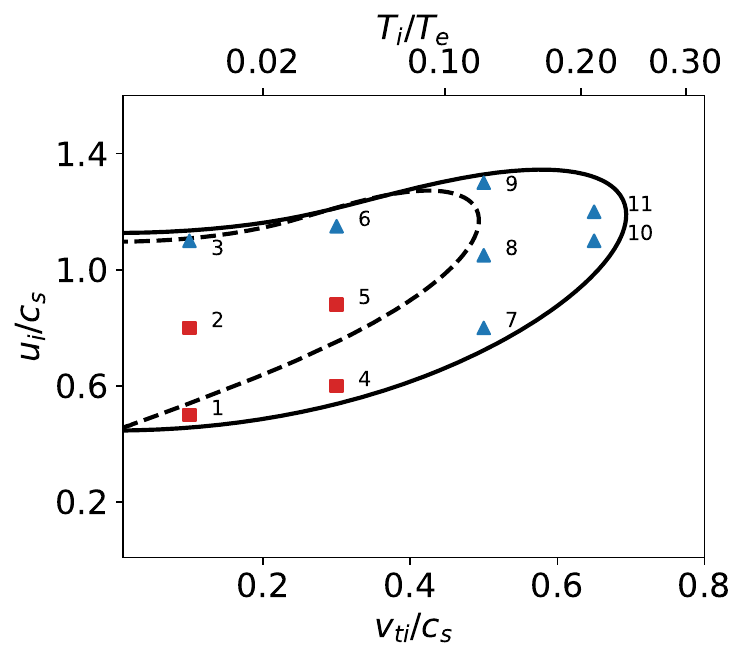}
    \caption{The parameters of $u_i$ and $v_{ti}$ tested in the simulations.
    The values of these parameters are listed in Table \ref{tab:sim-params}.
    The corresponding run numbers are labelled on the right side of the points.
    The solid and dashed lines are the same as those in Fig. \ref{fig:d2f-ts}.
    The solitary wave remains stable over the simulation time $t=500 \omega_{pe}^{-1}$ with the parameters denoted by the blue triangles.
    The instabilities occur for the parameters denoted by the red squares.
    }
    \label{fig:sim-ver}
\end{figure}

The initial ion distribution is set as the two-stream distribution \eqref{eq:fi2}, which might be a realistic scenario in space plasmas. \cite{Kamaletdinov2021}
\begin{figure*}[tb]
    \centering
    \includegraphics[width=15cm]{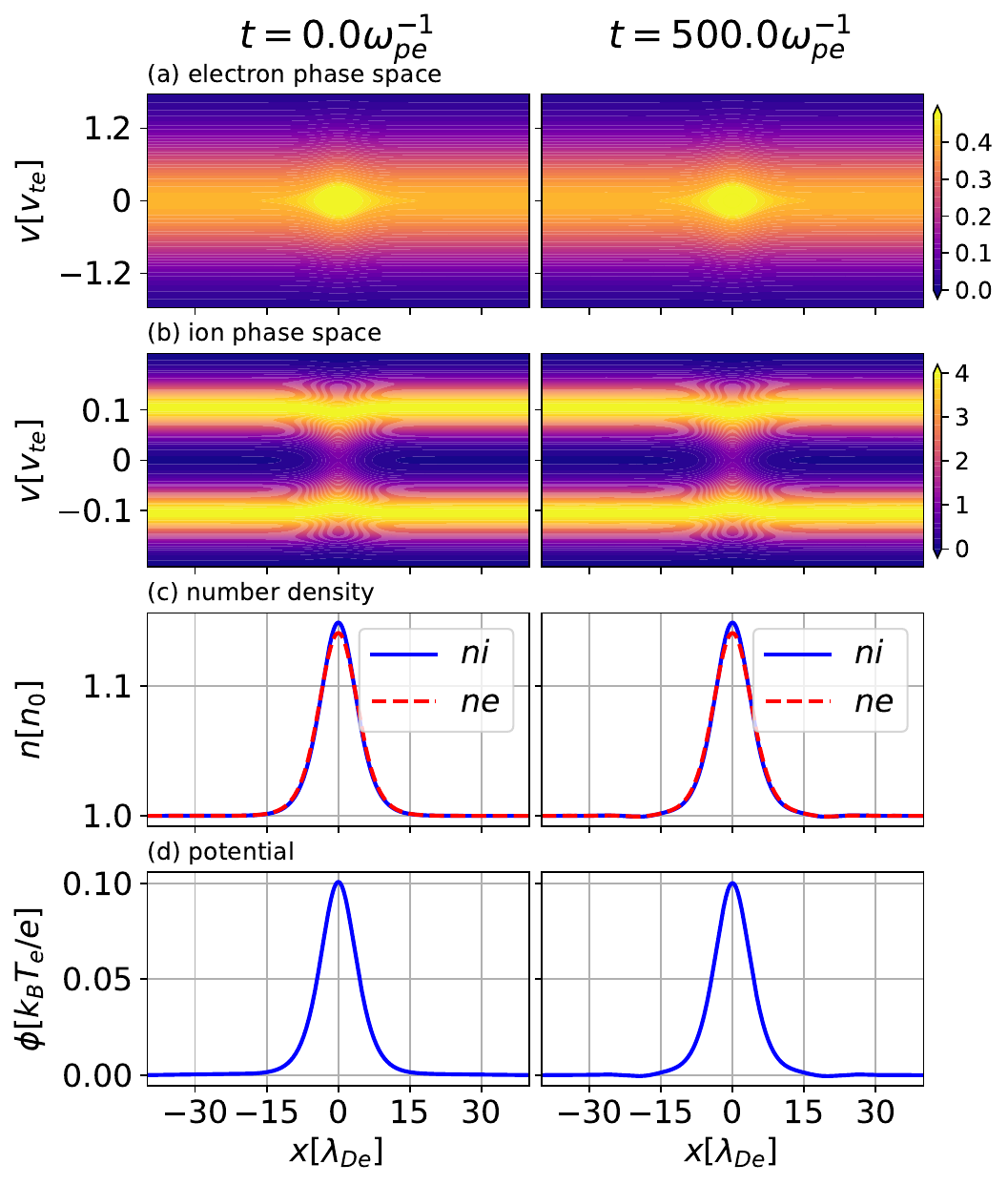}
    \caption{Results of the stable case (Run 8) in the simulations.
        The parameters are $u_i = 0.105$ and $v_{ti} = 0.05$.
    }
    \label{fig:sim-case-stable}
\end{figure*}
\begin{figure*}[tb]
    \centering
    \includegraphics[width=15cm]{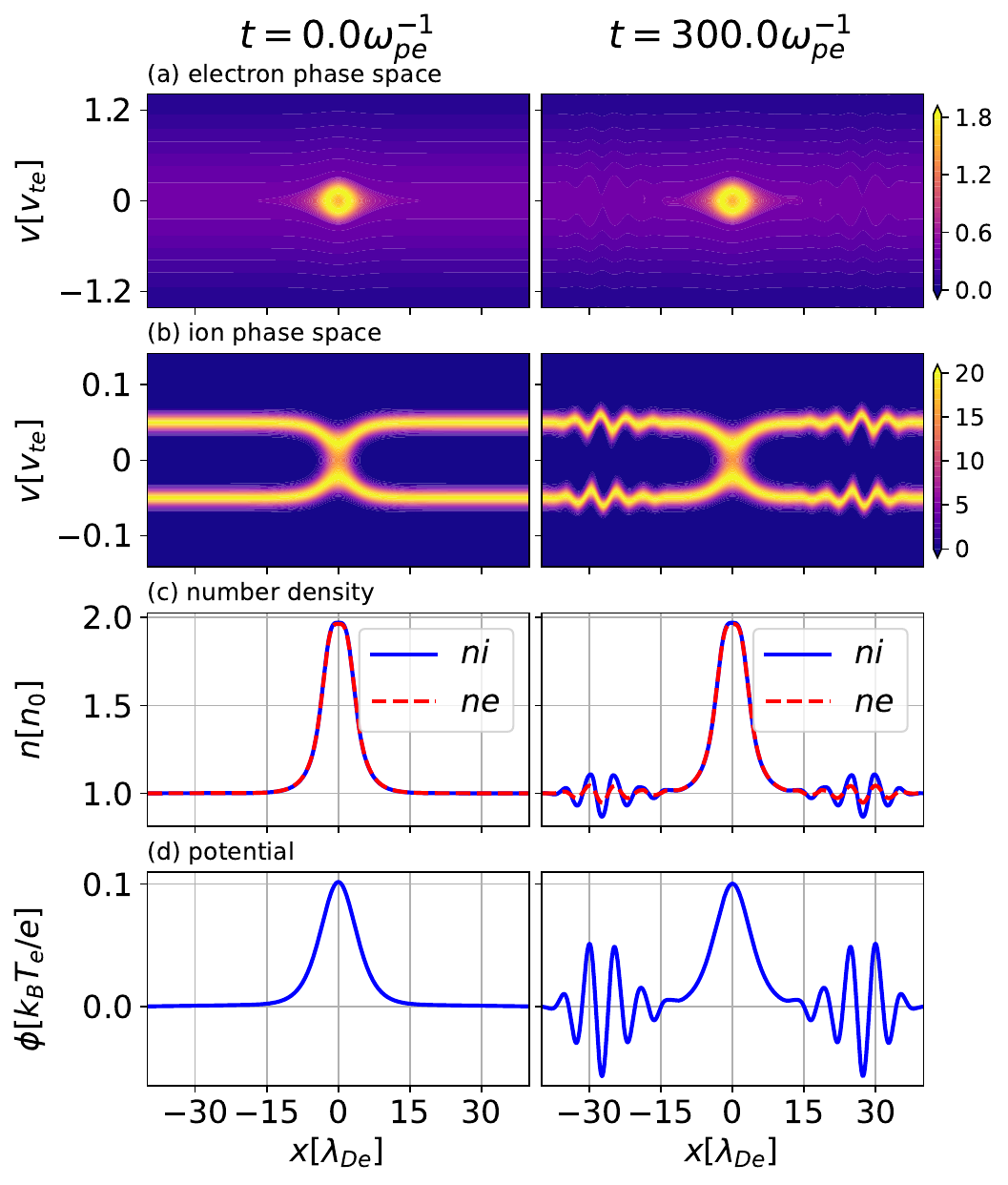}
    \caption{Results of the unstable case (Run 1) in the simulation.
        The parameters are $u_i = 0.05$ and $v_{ti} = 0.01$. 
    }
    \label{fig:sim-case-unstable}
\end{figure*}
The initial passing electron distribution is the Maxwellian one \eqref{eq:fp}, while the trapped one is the collection of \eqref{eq:ft1}, \eqref{eq:ft2}, and \eqref{eq:ft3-ts}.
The initial potential is set as \eqref{eq:phi}.
The test parameters of $u_i$ and $v_{ti}$ are shown in Fig. \ref{fig:sim-ver} and Table \ref{tab:sim-params}, including both the IAS with a local minimum and maximum at the center $x=0, v=0$. 
The other parameters are $\psi = 0.1$, $\Delta = 5$, and $\mu = 50$, resulting in the dimensionless ion-acoustic speed $c_s = 0.1$.
To the end of the simulation $t=500 \omega_{pe}^{-1}$, 
the solitary wave remains stable when we take the parameters represented by the blue triangles in Fig. \ref{fig:sim-ver}.
However, the instability occurs for the parameters represented by the red squares. 
Two examples of the stable and unstable cases are shown in Figs. \ref{fig:sim-case-stable} and \ref{fig:sim-case-unstable}, respectively.
\begin{table}[tb]
\begin{tabular}{ccccccc}
\hline
\hline
Run~~      & 1~~    & 2~~     & 3~~    & 4~~     & 5~~     & 6     \\
\hline
$v_{ti}$~~ & 0.01~~ & 0.01~~  & 0.01~~ & 0.03~~  & 0.03~~  & 0.03  \\
$u_i$~~    & 0.05~~ & 0.08~~  & 0.11~~ & 0.06~~  & 0.088~~ & 0.115 \\
\hline
\hline
Run~~      & 7~~    & 8~~     & 9~~    & 10~~    & 11~~    &       \\
\hline
$v_{ti}$~~ & 0.05~~ & 0.05~~  & 0.05~~ & 0.065~~ & 0.065~~ &       \\
$u_i$~~    & 0.08~~ & 0.105~~ & 0.13~~ & 0.11~~  & 0.12~~  &       \\
\hline
\hline
\end{tabular}
\caption{The parameter values in the simulations.}
\label{tab:sim-params}
\end{table}


This instability can be attributed to the background distribution, which can be confirmed by setting $\psi=0$ initially in simulations.
\begin{figure}[tb]
    \centering
    \includegraphics[width=8.5cm]{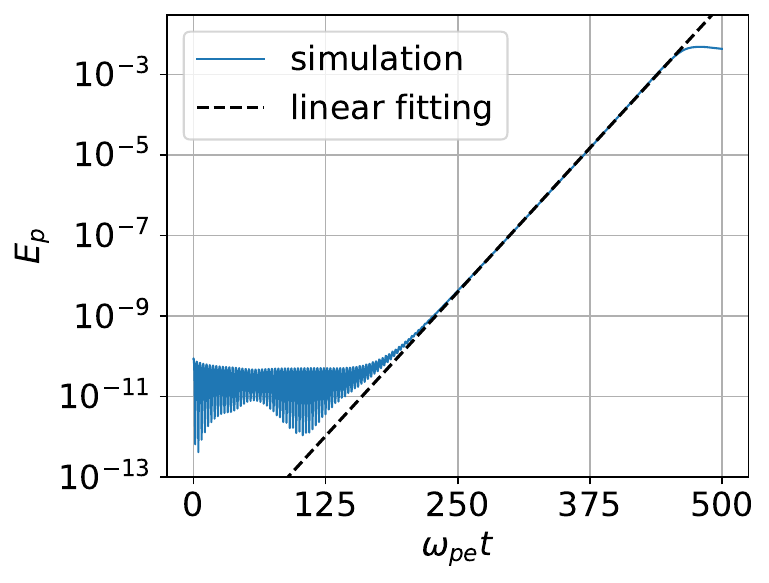}
    \caption{
        The evolution of the potential energy in the simulation using the Run 1 parameters, but with $\psi=0$ at the initial time.
        The dashed line denotes the linear fit of the growth rate, which is $\gamma/\omega_{pe} \simeq 0.033$.
    }
    \label{fig:sim-r1-psi0}
\end{figure}
Figure \ref{fig:sim-r1-psi0} illustrates the time evolution of the potential energy in the simulation by choosing the parameters of Run 1 with $\psi=0$.
The theoretical growth rate can be solved by the linear dispersion relation, \cite{Krall1973}
\begin{equation}
    1 + \sum_s \frac{1}{k^2 \lambda_{Ds}^2} \left[1+\zeta_s Z (\zeta_s) \right] =0,
    \label{eq:disp}
\end{equation}
where $s$ denotes the species, including the non-drifting electrons and the ions with the positive and negative drifting speed $\pm u_i$, respectively.
The numerical solution of Eq. \eqref{eq:disp} gives the maximum growth rate $\gamma/\omega_{pe} \simeq 0.035$ at $k \lambda_{De} \simeq 1.162$, which matches the growth rate $\gamma/\omega_{pe} \simeq 0.033$ calculated from the simulation.
Therefore, the instability observed in the simulation is a linear ion-ion instability triggered by the background distributions.
In addition, no acceleration of the solitary potential is observed during the simulations.
Consequently, our results could confirm that the constructed IAS model is stable when the background distribution is linearly stable.

\section{Summary and Discussion}
\label{sec:summary}
In this work, we study the solitary waves by the BGK integral method with the ion response.
Due to the fully kinetic model, the behaviors of the IAS are different from those derived by the well-known fluid method in the literature.
We consider two specific cases of ions, i.e., the single stream with the WB distribution and the two counter streams with the Maxwellian distribution.
For the case of the single ion stream, the trapped electron distribution \eqref{eq:ft} is derived,
and the result indicates that this trapped distribution can be either a hole or a hump in the phase space, representing the EH and IAS, respectively.
To distinguish between the EH and IAS, a critical condition \eqref{eq:cond-sep} is proposed, i.e., the trapped distribution at the center $w=-\psi$ is smaller/larger than that at the edge $w=0$.
The separatrices of the IAS and EH in parameter spaces are plotted in Fig. \ref{fig:uc}.
For the case of two counter-streaming Maxwellian ions, the trapped electron distribution is studied numerically, and the separatrices are shown in Fig. \ref{fig:sep_s_eh}.
The results indicate that the solitary wave with the amplitude $\psi \sim 0.1 k_B T_e/e$ and the width $\Delta \sim 10 \lambda_{De}$, usually observed in the space plasmas, could be recognized as the IAS only in the case of $T_i/T_e \lesssim 0.3$. 
At last, we perform Vlasov simulations to test the stability of the IAS model constructed in this work.
It shows the IAS constructed by the BGK method is stable if the background distributions are linearly stable.
Our conclusions may be applied to identify whether the solitary waves is an IAS or EH in the observations.

\begin{acknowledgments}
This work was supported by the National Natural Science Foundation of China (No. 12105361) and by the Scientific Research Startup Foundation of Civil Aviation University of China (No. 2015QD05X).
\end{acknowledgments}

\section*{Data Availability}
The data that support the findings of this study are available from the corresponding author upon reasonable request.

\appendix
\section{The consistency of the trapped electron distribution in the fluid soliton model}
\label{app:consistency}
The Sagdeev potential for the pure soliton with the cold ion assumption is given by, \cite{Davidson1972}
\begin{equation}
    V(\phi) = 1-e^\phi+M^2-M\sqrt{M^2-2\phi},
    \label{eq:V-soliton}
\end{equation}
in which $M$ is the Mach number.
Recalling that the speeds $u_i$ and $c_s$ are in the unit of $v_{te}$ in this work, one obtains the Mach number $M = u_i/c_s = \sqrt{2\mu}u_i$.
Substituting the Sagdeev potential \eqref{eq:V-soliton} into Eq. \eqref{eq:ft2-general}, one can obtain,
\begin{align}
    f_{e,t}^{(2)} 
    &= \frac{1}{\pi} \int_0^{-w} \left[
        e^\phi - \frac{u_i}{2\mu(u_i^2-\phi/\mu)^{3/2}}
        \right]\frac{\dd{\phi} }{\sqrt{-w-\phi}} \notag\\
    &= \frac{1}{\sqrt{\pi}}e^{-w}\erf(\sqrt{-w}) - \frac{\sqrt{-w}}{\pi(\mu u_i^2+w)}.
\label{eq:ft2_soliton} 
\end{align}
In the limit of $T_i \rightarrow 0$, Eq. \eqref{eq:ft3-os} yields,
\begin{align}
    f_{e,t}^{(3)} 
    &= \lim_{v_{ti}\rightarrow 0} \left[ \frac{1}{2\sqrt{\mu} v_{ti} \pi} \arctanh \left(\frac{2v_{ti}\sqrt{-w/\mu}}{u_i^2-v_{ti}^2+w/\mu}\right)\right] \notag\\
    &= \frac{\sqrt{-w}}{\pi(\mu u_i^2+w)},
    \label{eq:ft3-soliton}
\end{align}
by applying L'Hopital's rule.
Therefore, the total trapped electron distribution is,
\begin{equation}
    f_{e,t} = f_{e,t}^{(1)} + f_{e,t}^{(2)} + f_{e,t}^{(3)} = \frac{1}{\sqrt{\pi}}e^{-w},
    \label{eq:ft_soliton}
\end{equation}
which is exactly the Maxwellian distribution supposed in the fluid theory of the pure soliton.

\section{The sign of \texorpdfstring{$\pdv*[2]{f_e}{v}$}{∂²fe/∂v²} at \texorpdfstring{$x=0, v=0$}{x=0, v=0} without the ion response}
\label{app:sign}
The second partial derivative of the trapped electron distribution with respect to the velocity $v$ at the center $x=0, v=0$ reads,
\begin{equation}
    \pdv[2]{f_{e,t}}{v}\Big\rvert_{x=0,v=0} = \frac{2}{\pi\sqrt{\psi}}  - \frac{2e^\psi}{\sqrt{\pi}} \erfc(\sqrt{\psi}) 
                            + \frac{40}{\pi\Delta^2 \sqrt{\psi}},
    \label{eq:d2f-no-ions}
\end{equation}  
when the ions are assumed to be motionless and uniform spatially.
The last term is always positive.
The first two terms can be rewritten as,
\begin{equation}
    \frac{2}{\pi\sqrt{\psi}}  - \frac{2e^\psi}{\sqrt{\pi}} \erfc(\sqrt{\psi}) 
    = \frac{2e^\psi}{\pi} \left[\frac{e^{-\psi}}{\sqrt{\psi}} - \sqrt{\pi} \erfc(\sqrt{\psi})\right].
\end{equation}
The term in the square bracket is a monotonic decreasing function of $\psi$ for $\psi>0$, which could be proved by the negativity of its derivative,
\begin{equation}
    \dv{}{\psi}\left[\frac{e^{-\psi}}{\sqrt{\psi}} - \sqrt{\pi} \erfc(\sqrt{\psi})\right] = -\frac{e^{-\psi}}{2\psi^{3/2}}<0.
\end{equation}
Therefore, one has, for $\psi>0$,
\begin{equation}
    \frac{e^{-\psi}}{\sqrt{\psi}} - \sqrt{\pi} \erfc(\sqrt{\psi})>\left[\frac{e^{-\psi}}{\sqrt{\psi}} - \sqrt{\pi} \erfc(\sqrt{\psi})\right]_{\psi=\infty} = 0,
    \label{eq:ineq}
\end{equation}
resulting in the second partial derivative \eqref{eq:d2f-no-ions} being positive in the absence of ion response.

\section{The solutions of \texorpdfstring{$u_c^*$}{uc*}}
\label{app:sol-ucs}
The equation of $u_c^*$,
\begin{multline}
    \frac{2}{\pi\sqrt{\psi}}  - \frac{2e^\psi}{\sqrt{\pi}} \erfc(\sqrt{\psi}) + \frac{40}{\pi\Delta^2 \sqrt{\psi}} \\
    - \frac{1}{\mu\pi\sqrt{\psi}} \frac{u_i^2-v_{ti}^2+\psi/\mu}{(u_i^2-v_{ti}^2+\psi/\mu)^2-4v_{ti}^2\psi/\mu}
    =0,
\end{multline}
gives two solutions,
\begin{equation}
    u_{c1}^{*2} = v_{ti}^2 + \frac{\psi}{\mu} + \frac{1}{4 \mu A} 
                + \sqrt{\frac{1}{16 \mu^2 A^2} + \frac{\psi}{\mu^2 A} + 4 v_{ti}^2 \frac{\psi}{\mu}},
\end{equation}
\begin{equation}
    u_{c2}^{*2} = v_{ti}^2 + \frac{\psi}{\mu} + \frac{1}{4 \mu A} 
                - \sqrt{\frac{1}{16 \mu^2 A^2} + \frac{\psi}{\mu^2 A} + 4 v_{ti}^2 \frac{\psi}{\mu}}.
\end{equation}
Because $A = 1-\sqrt{\pi \psi}\exp(\psi)\erfc(\sqrt{\psi})+20/\Delta^2 > 0$ for positive $\psi$ due to Eq. \eqref{eq:ineq}, 
we have
\begin{equation}
    \frac{1}{4 \mu A} - \sqrt{\frac{1}{16 \mu^2 A^2} + \frac{\psi}{\mu^2 A} + 4 v_{ti}^2 \frac{\psi}{\mu}} <0
\end{equation}
resulting in $u_{c2}^{*2} < v_{ti}^2 + \psi/\mu < u_r^2$.
Consequently, the solution $u_{c2}^*$ is neglected.

\section{Derivations of \texorpdfstring{$\pdv*[2]{f_e}{v}$}{∂²fe/∂v²} at \texorpdfstring{$x=0, v=0$}{x=0, v=0} in the case of two counter-streaming Maxwellian ions}
\label{app:dev-d2f}
Recalling the electron energy $w = v^2-\phi$, one can calculate the partial derivative by,
\begin{equation}
    \pdv[2]{}{v} = 2\pdv{}{w} + 4v^2 \pdv[2]{}{w},
\end{equation} 
leading to
\begin{equation}
    \pdv[2]{f_{e,t}}{v}\Big|_{x=0,v=0} = 2\pdv{f_{e,t}}{w}\Big|_{x=0,v=0},
    \label{eq:d2fv-dfw}
\end{equation} 
if $\pdv*[2]{f_{e,t}}{w}$ is finite at $x=0, v=0$.
Therefore, by differentiating Eq. \eqref{eq:ft3-ts}, we derive
\begin{align}
    \pdv{f_{e,t}^{(3)}}{w} = &-\frac{1}{\pi^{3/2}\mu v_{ti}^3} \pdv{}{w} \int_0^{-w} \frac{I(\phi)\dd{\phi}}{\sqrt{-w-\phi}} \notag \\
                           = &\frac{1}{\pi^{3/2}\mu v_{ti}^3} \left[\frac{I(0)}{\sqrt{-w}} + \int_0^{-w} \frac{\dv*{I}{\phi}}{\sqrt{-w-\phi}}\dd{\phi}\right].
    \label{eq:d2f-dev}
\end{align}
Substituting into Eq. \eqref{eq:d2fv-dfw} with $w=-\psi$, one obtains the last term in Eq. \eqref{eq:d2f-ts}.
\bibliography{library}
\end{document}